\documentclass[pdflatex,sn-mathphys-ay,iicol]{sn-jnl}

\usepackage{graphicx}%
\usepackage{multirow}%
\usepackage{amsmath,amssymb,amsfonts}%
\usepackage{amsthm}%
\usepackage{mathrsfs}%
\usepackage[title]{appendix}%
\usepackage{xcolor}%
\usepackage{textcomp}%
\usepackage{manyfoot}%
\usepackage{booktabs}%
\usepackage{algorithm}%
\usepackage{algorithmicx}%
\usepackage{algpseudocodex}%

\usepackage{siunitx}
\usepackage{braket}
\usepackage{cleveref}
\crefname{section}{Section}{Sections}
\crefname{equation}{Eq.}{Eqs.}
\crefname{figure}{Fig.}{Figs.}
\crefname{table}{Table}{Tables}
\crefname{appendix}{Appendix}{Appendices}
\crefname{algorithm}{Algorithm}{Algorithm}
\usepackage{stfloats}

\usepackage{ftnright}
\usepackage{mathtools}

\usepackage{etoolbox}           
\newcommand{\B}{\fontseries{b}\selectfont} 
\robustify\B

\makeatletter
\expandafter\patchcmd\csname\string\algorithmic\endcsname{\itemsep\z@}{\itemsep=0.2ex}{}{}
\makeatother

\newcolumntype{L}[1]{>{\raggedright\let\newline\\\arraybackslash\hspace{0pt}}m{#1}}
\newcolumntype{C}[1]{>{\centering\let\newline\\\arraybackslash\hspace{0pt}}m{#1}}
\newcolumntype{R}[1]{>{\raggedleft\let\newline\\\arraybackslash\hspace{0pt}}m{#1}}

\theoremstyle{thmstyleone}%
\theoremstyle{thmstyletwo}%

\theoremstyle{thmstylethree}%

\begin{document}

\title{Quantum reinforcement learning in dynamic environments}

\author*[1]{\fnm{Oliver} \sur{Sefrin}}\email{oliver.sefrin@dlr.de}
\author[2]{\fnm{Manuel} \sur{Radons}\nomail}
\author[2]{\fnm{Lars} \sur{Simon}\nomail}
\author[1,3]{\fnm{Sabine} \sur{W\"olk}\nomail}

\affil[1]{\orgdiv{Institute of Quantum Technologies}, \orgname{German Aerospace Center (DLR)}, \orgaddress{\street{Wilhelm-Runge-Straße 10}, \city{Ulm}, \postcode{89081}, \country{Germany}}}
\affil[2]{\orgname{Bundesdruckerei GmbH}, \orgaddress{\street{Kommandantenstraße 18}, \city{Berlin}, \postcode{10969}, \country{Germany}}}
\affil[3]{\orgdiv{Center for Integrated Quantum Science
and Technology (IQST)}, \orgname{Ulm University}, \orgaddress{\city{Ulm}, \postcode{89081}, \country{Germany}}}

\abstract{
    Combining quantum computing techniques in the form of amplitude amplification with classical reinforcement learning has led to the so-called ``hybrid agent for quantum-accessible reinforcement learning”, which achieves a quadratic speedup in sample complexity for certain learning problems.
    So far, this hybrid agent has only been applied to stationary learning problems, that is, learning problems without any time dependency within components of the Markov decision process.
    
    In this work, we investigate the applicability of the hybrid agent to dynamic RL environments.
    To this end, we enhance the hybrid agent by introducing a dissipation mechanism and, with the resulting learning agent, perform an empirical comparison with a classical RL agent in an RL environment with a time-dependent reward function.
    Our findings suggest that the modified hybrid agent can adapt its behavior to changes in the environment quickly, leading to a higher average success probability compared to its classical counterpart.
}


\maketitle

\section{Introduction}
\label{sec:intro}

From the success of convolutional neural networks for computer vision tasks in the early 2010s~\citep{ciresan2011flexible,krizhevsky2017alexnet} to the nowadays ever-expanding usage of generative artificial intelligence based on diffusion~\citep{sohldickstein2015diffusion} or transformer~\citep{vaswani2017attention} models, the vast field of machine learning (ML) has impacted many areas of modern life.
Besides the continuous development of model architectures and training methods, the rapid improvement of specialized computing hardware in the form of graphics or tensor processing units (GPUs/TPUs) has led to today's prevalence of ML.
This work focuses on reinforcement learning (RL), the ML subcategory encompassing interaction-based learning.
While best known in the area of games with algorithmic breakthroughs such as deep Q-learning (DQN) for Atari games~\citep{mnih2015human} or the MuZero algorithm~\citep{schrittwieser2020muzero}, RL also serves other areas such as the training of the above mentioned language models through reinforcement learning from human feedback (RLHF)~\citep{ziegler2020rl-from-human-feedback}.

The combination of RL with quantum computing, which promises polynomial~\citep{grover1997quantum,hamann2022performance} or even exponential~\citep{shor1994factoring,dunjko2018exponential,liu2021riorous,molteni2024exponential} speedups in specific problem tasks, has opened up the research field of quantum reinforcement learning (QRL)~\citep{dong2008quantum,dunjko2016quantum,jerbi2021parametrized}.
Due to the restrictions of current quantum hardware in the so-called noisy, intermediate-scale quantum (NISQ)~\citep{preskill2018nisq} era, however, the search for ``quantum advantage" remains challenging~\citep{schuld2023quantum,cerezo2024does,ciliberto2020statistical}.
Here, the so-called ``hybrid agent for quantum-accessible reinforcement learning"~\citep{hamann2022performance}, or briefly hybrid learning agent, provides an interesting outlook for the post-NISQ era: by incorporating amplitude amplification into the process of sampling action sequences, it allows to find rewarded sequences quadratically faster for specific RL environment types.

This hybrid learning agent, however, has only been used in stationary environments so far, i.e., environments which do not undergo changes throughout the training process.
Yet, many RL problems, especially regarding real-world applications (e.g., traffic control), do possess some kind of time-dependent dynamics (e.g., closed lanes due to road maintenance or temporary differences in traffic volume due to daily commuters). 
In such scenarios, a well-trained agent often performs worse after a sudden change of the environment compared to an untrained or a barely trained agent. 
As a consequence, a fast learning quantum agent often has a disadvantage compared to a slower learning classical agent when the environment changes. 
To compensate for this disadvantage and outperform classical agents again, the hybrid agent needs an efficient procedure to unlearn or relearn.
Therefore, in this work, we take a first step by introducing a dissipation mechanism in the hybrid learning agent and by investigating its applicability and efficiency in such non-stationary environments. 
In this way, the hybrid learning agent becomes more robust to changes of the problem environment without the necessity to introduce complex mechanisms for explicitly detecting anomalies.
Specifically, we compare the hybrid learning agent against a classical counterpart in the so-called Gridworld or maze environment, where we add time dependency to the reward function both on the intra- and inter-episode level.
Our research questions are the following:
\begin{itemize}
    \item Which modifications have to be applied to the original hybrid learning agent as defined in~\citep{hamann2022performance} to achieve high-performance learning in these kinds of non-stationary environments?
    \item Assuming a better learning performance on a first reward function, can the hybrid learning agent overcome the resulting worse initial success probability after a change of reward function?
    \item Does the less frequent updating of the agent's policy due to the episodes spent on amplitude amplification hinder the hybrid learning agent's capability to adapt in the dynamic environment?
\end{itemize}

This article is structured as follows: in~\cref{sec:background}, we recapitulate RL theory and the initial hybrid learning agent and summarize related work on QRL and continual RL.
\Cref{sec:problem} features a description of the basic Gridworld problem as well as the modified environment options with non-stationary reward functions.
In~\cref{sec:methods}, we present our modifications of the hybrid learning agent for the non-stationary RL problems and discuss the methodology of our numerical simulations.
\cref{sec:results,sec:discussion} comprise our simulation results and the discussion thereof.
Finally in~\cref{sec:conclusion}, we provide a conclusion and an outlook to future research.

\section{Background}
\label{sec:background}

\subsection{Reinforcement Learning}
\label{sec:background:sub:rl}

Reinforcement Learning (RL)~\citep{sutton2018reinforcement} is a branch of machine learning focused on learning through interaction.
Usually, this interaction takes place in a sequence of discrete time steps $t\in\mathbb{N}_0$ and is formalized as a \textit{Markov Decision Process} (MDP)~\citep{puterman1994mdp}.
The MDP is a five-tuple 
\begin{align}\label{eq:mdp}
    M=\langle \,T, \,\mathcal{S}, \,\mathcal{A}, \,p_t(\cdot | s,a), \,r_t(s,a) \,\rangle.
\end{align}
$T$ refers to the total amount of interaction steps and can be finite (in \textit{episodic} tasks) or infinite (in \textit{continuing} tasks).
$\mathcal{S}$ and $\mathcal{A}$ denote the sets of states and actions, respectively.
The \textit{transition probability function} 
\begin{align}\label{eq:transition-prob-func}
    p_t : \mathcal{S} \times \mathcal{S} \times \mathcal{A} \rightarrow [0,1]
\end{align} 
denotes the probability $p_t(s'| s,a)$ that upon playing action $a\in\mathcal{A}$ in state $s\in\mathcal{S}$ at time $t$, the state at time $t+1$ will be $s'\in\mathcal{S}$.
Associated with this new state is a real-valued, bounded reward $R_{t+1}$, with the value determined by the \textit{reward function}
\begin{align}\label{eq:reward-func}
    r_t:\mathcal{S} \times \mathcal{A} \rightarrow \mathbb{R}.
\end{align} 
After playing action $a$ in state $s$ at time $t$, the agent receives the reward $R_{t+1}:=r_t(s,a)$.

The agent's behavior is described by its \textit{policy} $$\pi:\mathcal{A} \times \mathcal{S} \rightarrow [0,1], $$ which describes the probability $\pi(a|s)$ to choose action $a$ upon perceiving state $s$.
The agent's goal is to adapt its policy such as to maximize the expected \textit{cumulative reward}
\begin{align}
    G_t = \sum_{k=t+1}^T \chi^{k-t-1} R_{k}
\end{align}
at any time step $t$, where $\chi\in[0,1]$ is a discount factor.
In continuing tasks, setting $\chi < 1$ ensures that the cumulative reward does not diverge. 
In episodic tasks, the interaction terminates once the final time step $t=T$ is reached, upon which the interaction starts again at $t=0$ from an initial state $s_0$.

The environment can be classified as \textit{stationary} (\textit{non-stationary/dynamical}) if the transition dynamics of the MDP are time-independent (time-dependent).
In the stationary case, we may omit the index $t$ from the transition probability function (\cref{eq:transition-prob-func}) and the reward function (\cref{eq:reward-func}).
RL in non-stationary environments can be further distinguished according to characteristics of the time dependency and the amount of information given to the agent about changes in the environment~\citep{padakandla2021survey-rl-dynamic-envs}.
Relevant to this work is \textit{continual RL} as defined in~\citep{ring1994continual,abel2023continual}, which describes an ongoing learning process in non-stationary environments where the ``best agents never stop learning"~\citep{abel2023continual}.
An example for this is an ensemble of MDPs from which an MDP is randomly sampled repeatedly after certain amounts of time steps, such that an agent must constantly adapt to the current MDP.

The speedup of the hybrid learning agent, as defined in the following section, has been proven for a special class of RL tasks, so-called \textit{deterministic and strictly episodic} (DSE) tasks (or environments).
``Strictly episodic'' signifies that the episode length $T$ is not only finite, but also has a fixed value for all episodes.
``Deterministic'' refers to the transition probability function of~\cref{eq:transition-prob-func}, which, in the deterministic case, returns for any state-action tuple $(s,a)$ unity for one specific state $s'$ and zero for all other states.
Thus, we can define the deterministic version of $p_t$, 
\begin{align}\label{eq:transition-prob-func-deterministic}
    p_t^{(\mathrm{det})}: \mathcal{S} \times \mathcal{A} \rightarrow \mathcal{S},
\end{align}
which directly projects to the next state $s'$.

\subsection{Hybrid Learning Agent}
\label{sec:background:sub:hybrid-agent}

The ``hybrid agent for quantum-accessible reinforcement learning"~\citep{hamann2022performance}, which we further refer to as the hybrid learning agent, is a hybrid quantum-classical algorithm that can be used in combination with classical RL algorithms to provide a quasi quadratic speedup in sample complexity.
In brief, in its quantum component, the hybrid learning agent uses amplitude amplification~(AA) to increase the probability to sample rewarded action sequences.
This sampled action sequence is subsequently tested in an episode of classical interaction to determine the sequence of encountered percepts and the reward information, which can then be used to update the agent's policy.

In more detail, the algorithm operates in a quantum communication scenario between the agent and the environment.
First, assuming finite action, percept, and reward spaces $\mathcal{A}$, $\mathcal{S}$, and $\mathcal{R}\subset\mathbb{R}$, we use the following encoding:
\begin{itemize}
    \item all actions $a\in\mathcal{A}$ are encoded in a set of orthonormal quantum states $\ket{a}_A$ in Hilbert space $\mathcal{H}_A$,
    \item all percepts $s\in\mathcal{S}$ are encoded in a set of orthonormal states $\ket{s}_S$ in Hilbert space $\mathcal{H}_S$,
    \item and all rewards $r\in\mathcal{R}$ are encoded in a set of orthonormal states $\ket{r}_R$ in Hilbert space $\mathcal{H}_R$.
\end{itemize}

The agent sends quantum states $\ket{a}_A$ to the environment and receives quantum states $\ket{s}_S$ and $\ket{r}_R$, accordingly.
From here on, we consider DSE environments with a binary episode reward $r$.
That is, exactly one interaction step in an episode may receive a reward of one; re-visiting the rewarded percept later in the episode does not yield another reward.\footnote{Strictly speaking, this \textit{first-visit reward} violates the Markov property. Adding a first-visit flag to each percept remedies this formal violation.}
Further, we use vector notation $\Vec{a} = (a_0, a_1, \dots, a_{T-1})$ to denote a sequence of actions played in an episode or $\Vec{s} = (s_0, s_1, \dots, s_{T})$ for the sequence of percepts encountered in an episode.
As a short-hand notation for the sequence of encountered percepts and the reward outcome of playing action sequence $\Vec{a}$ after the initial state $s_0$, we write $\Vec{s}(s_0, \vec{a})$ and $r(s_0, \Vec{a})\in\{0,1\}$, respectively.
Lastly, let 
\begin{align}\label{eq:initial-percept-state}
    \ket{\vec{s}_\mathrm{init}}_S = \ket{(s_0,\varnothing,\dots,\varnothing)}_S \in \mathcal{H}_S^{\otimes (T+1)}    
\end{align}
denote the initial percept quantum state before the interaction with a fiducial state $\ket{\varnothing}_S$.

We can model the response of the environment to an episode-long sequence of actions $\Vec{a}$ as a unitary $U_\mathrm{env}$ acting on the multipartite state $\ket{\Vec{a}}_A \ket{\vec{s}_\mathrm{init}}_S \ket{0}_R$~\citep{dunjko2016quantum}:
\begin{align}\label{eq:env-unitary}
    \begin{split}
    &U_\mathrm{env} \ket{\Vec{a}}_A \ket{\vec{s}_\mathrm{init}}_S \ket{0}_R \\
    = &\ket{\Vec{a}}_A \ket{\Vec{s}(s_0, \Vec{a})}_S \ket{r(s_0,\Vec{a})}_R.
    \end{split}
\end{align}
As shown in~\citep{dunjko2016quantum}, with one query of $U_\mathrm{env}$ for simple environments with action-independent percepts (\textit{memoryless} environments) or two queries for regular DSE environments, one can create a phase-kickback oracle $O_\mathrm{env}$:
\begin{align}\label{eq:oracle}
    \begin{split}
    O_\mathrm{env} &\ket{\Vec{a}}_A \ket{\vec{s}_\mathrm{init}}_S \ket{-}_R \\
    = (-1)^{r(s_0,\Vec{a})} &\ket{\Vec{a}}_A \ket{\vec{s}_\mathrm{init}}_S \ket{-}_R.
    \end{split}
\end{align}
Using the current policy $\pi$ and, e.g., a mapping mechanism (see Appendix~\ref{app:sec:projective-simulation}), one can compute the probability to choose action sequence $\vec{a}$ under said policy, which we denote $\hat{\pi}(\vec{a})$.
This allows preparing the weighted superposition\footnote{Preparing the state by explicitly computing the sampling probability of each action sequence $\Vec{a}$ quickly becomes infeasible for larger sequence lengths due to the exponential scaling of the number of possible sequences. We do note, however, that the former can be achieved efficiently, e.g., by using quantum policies based on variational quantum circuits.}
\begin{align}\label{eq:initial-state}
    \ket{\psi} = \sum\limits_{\Vec{a}\in \mathcal{A}^{\otimes T}} \sqrt{\hat{\pi}(\Vec{a})} \ket{\Vec{a}}_A \ket{\vec{s}_\mathrm{init}}_S \ket{-}_R
\end{align}
and defining the diffusion operator 
\begin{align}\label{eq:diffusion-op}
    D=\mathbb{I} - 2 \ket{\psi}\bra{\psi}.
\end{align}
With the Grover operator $G = D O_\mathrm{env}$ we may thus perform AA~\citep{grover1997quantum,brassard2002quantum}.
By denoting the current classical success probability 
\begin{align}
    Q = \sum_{\{\Vec{a}| r(\Vec{a}) = 1\}} \hat{\pi}(\Vec{a}),
\end{align} 
we arrive at the probability of measuring a rewarded action sequence after $k$ iterations of AA,
\begin{align}\label{eq:grover-prob}
    p_{AA}(Q, k) = \sin^2\left(\left[2k+1\right] \arcsin(\sqrt{Q})\right).
\end{align}
The measurement gives a candidate action sequence $\vec{a}$, but reveals no percept or reward information due to the necessary uncomputation.
Hence, an episode of classical play following the quantum part is required to determine $\vec{s}(\vec{a})$ and $r(\vec{a})$. 
Consecutively, the learning agent uses this information to update its policy according to some update mechanism as  specified, e.g., in Projective Simulation (PS)~\citep{briegel2012projective}, Q-learning~\citep{watkins1992q}, REINFORCE~\citep{williams1992reinforce}, or quantum natural policy gradient~\citep{meyer2023quantum-natural-policy-gradients}.

Since the current success probability $Q$ is typically not known in RL, the hybrid learning agent uses the AA variation for an unknown number of solutions presented in~\citep{boyer1998tight}.
This requires a lower bound estimate $Q_\mathrm{min}$ of $Q$.

The full hybrid learning agent in a version agnostic to the specific policy update mechanism is presented in~\Cref{alg:hybrid-standard}.
In this notation, the policy update mechanism is specified in the
subroutine \texttt{policy\_update} (see~\Cref{alg:ps-policy-update} in Appendix~\ref{app:sec:projective-simulation} for an examplary mechanism using PS), 
which uses the interaction history of an episode to update the current policy.
Here, we update the policy just on positive feedback (i.e., if the chosen action sequence is rewarded), which is typical for RL with the PS algorithm.
However, an update with every type of feedback after each iteration is possible by simply moving lines~\ref{alg:hybrid-standard:line:policy-update} and~\ref{alg:hybrid-standard:line:q_min-update} of the algorithm into the while-loop and adding the reward information to the arguments of \texttt{policy\_update}.

\begin{algorithm}[!tb]
\caption{Hybrid Learning Agent of Hamann\,\&\,Wölk~\citep{hamann2022performance}}\label{alg:hybrid-standard}
\begin{algorithmic}[1]
    \vspace{3pt}
    \Require MDP $\langle \,T, \,\mathcal{S}, \,\mathcal{A}, \,p_t^{(\mathrm{det})}(s,a), \,r_t(s,a) \,\rangle$, initial percept $s_0$, \\Grover operator $G$, \\$\texttt{policy\_update}$ subroutine
    \State initialize algorithm constant $\lambda \gets 5/4$ 
    \State initialize lower bound estimate $Q_\mathrm{min}$ \label{alg:hybrid-standard:line:q_min-init}
    \State algorithm parameter $m \gets 1$ \label{alg:hybrid-standard:line:loop-start}
    \State $\texttt{rewarded} \gets \texttt{false}$
    \While{\texttt{not rewarded}}
        \LComment{Quantum Part}
        \State $k \gets \mathrm{random \;integer \; in \;[0,m)}$\label{alg:hybrid-standard:line:k}
        \State prep. $\ket{\psi} \gets \!\sum\limits_{\Vec{a}\in \mathcal{A}^{\otimes T}}\! \sqrt{\hat{\pi}(\Vec{a})} \ket{\Vec{a}}_A \ket{\vec{s}_\mathrm{init}}_S \ket{-}_R$
        \State amplitude amplification $\ket{\psi'} \gets G^k \ket{\psi}$ 
        \State $\Vec{a}'\gets$ \textbf{measure}  $\ket{\psi'}$
        \LComment{Classical Part}
        \For{$i=1$ \textbf{to} $T$}
            \State $s_i \gets p_i^{(\mathrm{det})}(s_{i-1}, a_{i-1})$
            \State $r_i \gets r_i(s_{i-1},a_{i-1})$
            \If{$r_i = 1$}
                \State \texttt{rewarded} $\gets$ \texttt{true}
                \State $\Vec{s} \gets (s_0, s_1, \dots, s_i)$
                \State \texttt{break}
            \EndIf
        \EndFor
        \State $m \gets \min\left( \lambda \cdot m, \:Q_\mathrm{min}^{-1/2}  \right)$\label{alg:hybrid-standard:line:m}
    \EndWhile
    \LComment{Classical RL Update}
    \State update $\pi$ using subroutine $\texttt{policy\_update}(\Vec{a}', \Vec{s})$ \label{alg:hybrid-standard:line:policy-update}
    \State update estimate $Q_\mathrm{min}$ \label{alg:hybrid-standard:line:q_min-update}
    \State repeat from step~\ref{alg:hybrid-standard:line:loop-start}
\end{algorithmic}
\end{algorithm}

\subsection{Related Work}
\label{sec:background:related-work}

The predominant category of QRL in the current NISQ era combines classical RL with so-called variational quantum circuits (VQCs)~\citep{cerezo2021variational}.
In these algorithms, a quantum circuit with trainable, parameterized gates takes the place of an artificial neural network as the estimator for, e.g., the state-value function in value-based RL such as the DQN algorithm~\citep{mnih2015human} or the policy in policy-gradient-based RL such as the DDPG algorithm~\citep{lillicrap2019continuous}.
The feasibility of VQC-powered QRL algorithms has been tested in Gymnasium~\citep{kwiatkowski2024gymnasium} tasks~\citep{jerbi2021parametrized,skolik2022quantum,chen2020variational,lockwood2020reinforcement}, standard maze problems~\citep{chen2024deep}, Atari games~\citep{lockwood2021playing}, and quantum state preparation problems~\citep{wu2025quantum}. 
In general, QRL approaches based on VQCs are able to learn to solve quantum as well as classical problems and can be implemented on existing NISQ devices. 
However, if and how it is in general possible to generate quantum advantages in this way is an important question which is still under scientific examination~\citep{gilfuster2025relation}.

QRL approaches which do not use VQCs typically aim at generating speedups via quantum sub-routines such as quantum mean estimation~\citep{ganguly2023quantum,zhong2024provably}, quantum linear system solvers~\citep{cherrat2023quantum}, or a combination of quantum phase estimation and Grover search~\citep{wiedemann2023quantum}.
Other methods involve a quantum random walk approach to a variant of the Gridworld problem~\citep{pozza2022quantum} or quantum sampling methods in energy-based RL~\citep{jerbi2021quantum-enhancements}. 
Usually, these approaches are harder to implement on real quantum computer hardware but exhibit a provable quantum advantage.

The speed-up in learning time of the hybrid learning agent has been proven~\citep{hamann2022performance} and demonstrated in a proof-of-principle experiment using a nanophotonic processor~\citep{saggio2021experimental}.
Further research on the hybrid learning agent has focused on the effect of changing oracles during the learning process~\citep{hamann2021quantum} and on developing episode length selection strategies for not strictly episodic learning problems~\citep{sefrin2025hybrid}.
Similar to the hybrid learning agent, a QRL algorithm using amplitude amplification based on quantum random walks has been developed~\citep{paparo2014quantum} and experimentally verified on an ion trap device~\citep{sriarunothai2018speeding}.

Continual RL is also investigated in classical machine learning and classical approaches can be further classified according to the type of non-stationarity and the learning goals present in the problem.
Learners which aim to adapt to changes in the MDP but at the same time retain good performance on previous tasks and thus avoid catastrophic forgetting~\citep{mccloskey1989catastrophic} typically employ techniques such as context detection~\citep{dasilva2006context-detection} or meta-learning~\citep{finn2017maml}.
This is in contrast to our learning algorithm, which actively employs a forgetting mechanism to learn the current task more quickly.
For an overview of continual RL, its subtopics, and relevant approaches, we refer the reader to~\citep{khetarpal2022towards}.

\section{Problem Setting}
\label{sec:problem}
The hybrid learning agent, as described in~\cref{sec:background:sub:hybrid-agent}, provides a quasi quadratic speedup in sample complexity for RL tasks which have deterministic and strictly episodic (DSE) environments.
A standard example for a DSE environment is the so-called \textit{Gridworld} or maze problem, which is described in~\cref{sec:problem:sub:gw}.
So far, however, the hybrid learning agent has only been used for stationary environments.
In this work, we investigate the applicability of the hybrid agent to non-stationary environments.
Two classes of such non-stationary environments, exemplified as variations of the textbook maze problem, are presented in~\cref{sec:problem:sub:dynamic}.

\subsection{Gridworld}
\label{sec:problem:sub:gw}
The Gridworld problem, in its most common form, is an environment consisting of cell states in a two-dimensional grid.
In an episode, the agent starts from a fixed starting position and tries to reach a specified target cell.
The set of actions are the four basic move directions $\mathcal{A}=\{\texttt{up}, \,\texttt{down}, \,\texttt{left}, \,\texttt{right} \}$, the set of states $\mathcal{S}$ consists of all cells in the grid.
The response of the environment to choosing a move direction is deterministic and time-independent: the agent's new position $s_{t+1}$ is always one step away in the move direction from the old position $s_t$.
Some of the Gridworld's cells may consist of walls, however.
In that case, if the agent were to move into a wall cell, it remains at its old position.
The agent receives a reward of one for reaching the target cell for the first time in the episode and a reward of zero for any other outcome.
Typically, the episode concludes as soon as the agent reaches the target cell for the first time.
This, however, results in a variable episode length, as the agent may not always require the same number of steps to reach the target while learning.
By imposing a fixed number of interaction steps for the episode, the Gridworld environment can be turned into a strictly episodic one. 
We will use this setting throughout this paper in order to concentrate on effects caused by time dependency of the learning problem. 
However, the hybrid agent can also deal with learning problems with variable episode length as investigated in~\citep{sefrin2025hybrid}.

As such, the Gridworld problem is an example for a stationary environment since both the transition probability function and the reward function are time-independent.
This means that the agent does not need to adapt to any changes during the learning.

\subsection{Non-Stationary Reward Environments}
\label{sec:problem:sub:dynamic}
In this work, we are interested in the hybrid learning agent's capability to adapt to RL tasks with changing, i.e., time-dependent rewards.
Especially, we want to investigate scenarios in which the hybrid learning agent's increased learning speed with an initial reward distribution is detrimental to its successive performance in a varied reward setting.
This can be the case, e.g., in RL tasks with an initial set of rewarded action sequences $\mathcal{A}_1$ which is, at some point, exchanged with a disjoint set of rewarded action sequences $\mathcal{A}_2$.
In such a case, training the policy to sample an element of $\mathcal{A}_1$ with high probability directly lessens the probability to sample an element from $\mathcal{A}_2$.

As our testbed for these investigations, we modify the Gridworld problem with fixed episode length from Sec.~\ref{sec:problem:sub:gw}.
In particular, we first introduce as a first step an explicit time-dependent reward by replacing the fixed target with a moving target.
That is, the rewarded state follows a fixed route and moves one step along that route synchronously to each step of the agent.
As in the standard Gridworld setting, the agent is rewarded upon encountering the rewarded state for the first time within the episode.
If the agent has not found the rewarded state at the end of the route, the episode ends without a reward and both the agent and the reward are reset to their initial cells.
Thus, the reward route length $L$ determines the episode length.
Sometimes, it might be disadvantageous for the agent to move from its current state because the reward state would move onto its current position.
To accommodate for this fact, we expand the set of basic actions by the non-action, i.e., $\mathcal{A}=\{\texttt{left},\, \texttt{right}, \,\texttt{up}, \,\texttt{down}, \,\texttt{do\:nothing} \}$.

\begin{figure}[!t]
    \centering
    \includegraphics[width=0.7\columnwidth]{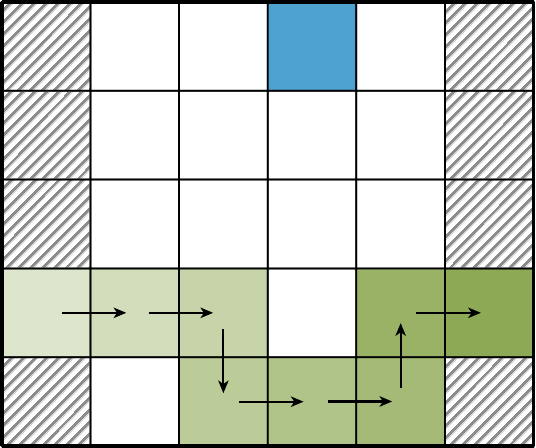}
    \caption{Graphical depiction of Gridworld layout~A, which is used for the fixed reward path scenario.
    The blue cell represents the agent's starting position, green cells indicate the reward path, which direction is indicated by arrows.
    Gray-shaded cells are impassable for the agent.}
    \label{fig:layout_a_vis}
\end{figure}

In a second step, we change the path of the moving target at a certain point during the learning process. 
This leads to the two learning scenarios which we investigate in this article:

\begin{itemize}
    \item[\textbf{A}] \textbf{Fixed Reward Path}\\
    This is the basic moving target scenario as introduced in the first step above.
    Here, the non-stationarity is purely restricted to the reward response within the episode.
    Across episodes, the reward response remains the same.
    A depiction of this Gridworld layout is given in Fig.~\ref{fig:layout_a_vis}.
    We assume that the agent has knowledge about the fact that the reward path is fixed, rendering a forgetting mechanism obsolete in this scenario.
    Using the hybrid learning agent in its standard form, we can use this scenario to verify its speedup compared to its purely classical counterpart. 
    \item[\textbf{B}] \textbf{Changing Reward Path}\\
    In this scenario, we start in the same fashion as in scenario A.
    However, after some time, we exchange the previous reward path for a new one with the same length $L$.
    This adds non-stationarity on the inter-episode level, as the reward response additionally changes after some amount of episodes.
    Therefore, this type of RL environment can be categorized as a hidden-mode Markov decision process (HM-MDP)~\citep{choi1999hidden}, where the different reward paths correspond to the different modes of the HM-MDP.
    The agent knows that a reward path switch may happen, but it is not notified about when it happens.
    Thus, using a forgetting mechanism is necessary in this scenario.
    In this example problem, we use two different reward paths and switch from the first to the second path upon triggering a pre-defined criterion.
    With this scenario, we challenge the hybrid learning agent's capability to relearn given a new reward situation.
\end{itemize}

\section{Methods}
\label{sec:methods}

In general, there are numerous methods to deal with changes in a learning problem. 
One option is to actively search for changes and then, e.g., reset the learning agent to its untrained starting status. 
This might be a good method if extensive changes are expected. 
More commonly, however, changes will be seldom and only of minor character. 
In such cases, a learning agent which is able to gradually adapt its behavior is more suited. 
Such a learning agent will be in general more robust since it neither sticks to a learned behavior nor inclines to sudden behavior changes. 
The dissipation mechanism of Projective Simulation (PS)~\citep{briegel2012projective} enables such a gradual adaptation. 
In~\cref{sec:methods:sub:forgetting-hybrid}, we introduce several modifications of our hybrid learning agent in order to use the dissipation mechanism of PS and capture changes in the MDP reasonably quickly.
Subsequently, in~\cref{sec:methods:sub:simulation}, we present the simulation setup used to compare the adapted hybrid learning agent versus a purely classical RL algorithm in non-stationary environments.

\subsection{Forgetting Mechanism for the Hybrid Learning Agent}
\label{sec:methods:sub:forgetting-hybrid}
Standard RL algorithms such as Q-Learning or SARSA are guaranteed to converge to the optimal behavior in stationary environments in the infinite time limit given sufficient exploration (infinite visits of all state-action pairs) and with a policy that converges to the greedy policy in that limit~\citep{sutton2018reinforcement}.
In non-stationary environments, however, these algorithms tend to capture changes in the MDP dynamics poorly~\citep{pieters2016q-learning,padakandla2021survey-rl-dynamic-envs}.
Using Q-Learning in continual RL, for example, requires a policy which enables exploration even in the infinite time limit such as an $\epsilon$-greedy policy with non-vanishing $\epsilon$~\citep{abel2023continual}.

In this work, we pair the hybrid learning agent with a policy-update mechanism based on PS, which is a tabular, state-action-value based RL algorithm.
In PS, a dissipation mechanism can be used to let the agent effectively adapt to changes in the environment and to enable more stable learning.
This dissipation drives the policy towards a uniform distribution -- therefore enhancing exploration -- with the strength of the dissipation governed by a parameter $\gamma\in[0,1]$.
For an overview of PS, the dissipation mechanism, and its relevant extensions to our work, see Appendix~\ref{app:sec:projective-simulation}.

To adapt the hybrid learning agent to non-stationary reward environments, we make two modifications.
The first modification targets the lower frequency of policy updates of the hybrid learning agent: 
To execute $k$ iterations of amplitude amplification, the hybrid learning agent requires $2k+1$ RL episodes.
Thus, it can only perform a policy update after $2k+1$ episodes, whereas a classical agent may update its policy after every episode.
To ensure that the hybrid learning agent receives the same amount of dissipation than a classical agent, we increase the dissipation strength according to the number of ``missed" update episodes.
The adapted update rule for the hybrid learning agent with a equivalency proof is shown in Appendix~\ref{app:sec:projective-simulation:sub:repeated-diffusion}.
The second modification is related to the lower bound estimate $Q_\mathrm{min}$ that the hybrid learning agent requires (cf. line~\ref{alg:hybrid-standard:line:m} of~\cref{alg:hybrid-standard}).
In general, this estimation can be performed in many different ways, such as (i) determining an absolute lower bound $Q_\mathrm{min}$ based solely on general problem settings, (ii) estimating $Q_\mathrm{min}$ based on observed rewards, or (iii) estimating $Q_\mathrm{min}$ based on the agent's current policy and observed rewards. 
For our simulations, we used option (iii) as described in detail in Appendix~\ref{app:sec:probability-estimation}.
To account for the possibility of changing rewards, we purge the set of found reward paths from the ones which are found to be no longer rewarded, which is further described in Appendix~\ref{app:sec:probability-estimation:sub:adaption}.

\subsection{Simulation Methodology}
\label{sec:methods:sub:simulation}
We compare the performance on the Gridworld scenarios stated above between the hybrid learning agent paired with PS based policy updates and classical PS.
For both the classical and the hybrid agent, we use the same PS specifications and hyperparameters.
These are the softmax-based policy of~\cref{eq:softmax-policy} with a pseudo-temperature of $\beta=1$, the glow extension with a fixed value of $\eta=0.05$, and the ECM with mapping. 
As dissipation parameter values, we test $\gamma \in \{0.01, 0.02, 0.05, 0.1\}$.

\subsubsection*{Gridworld Layouts}
To test scenario A, the fixed reward path scenario (cf.~\cref{sec:problem:sub:dynamic}), we use the Gridworld layout shown in~\cref{fig:layout_a_vis}, which we denote layout A.
With an episode length of seven, $5^7$~unique action sequences may be chosen by the agent.
\num{1330}~of these action sequences have a rewarded outcome, such that the initial success probability with a uniform policy is roughly~$\SI{1.7}{\percent}$.

\begin{figure}[!b]
    \centering
    \includegraphics[width=0.95\columnwidth]{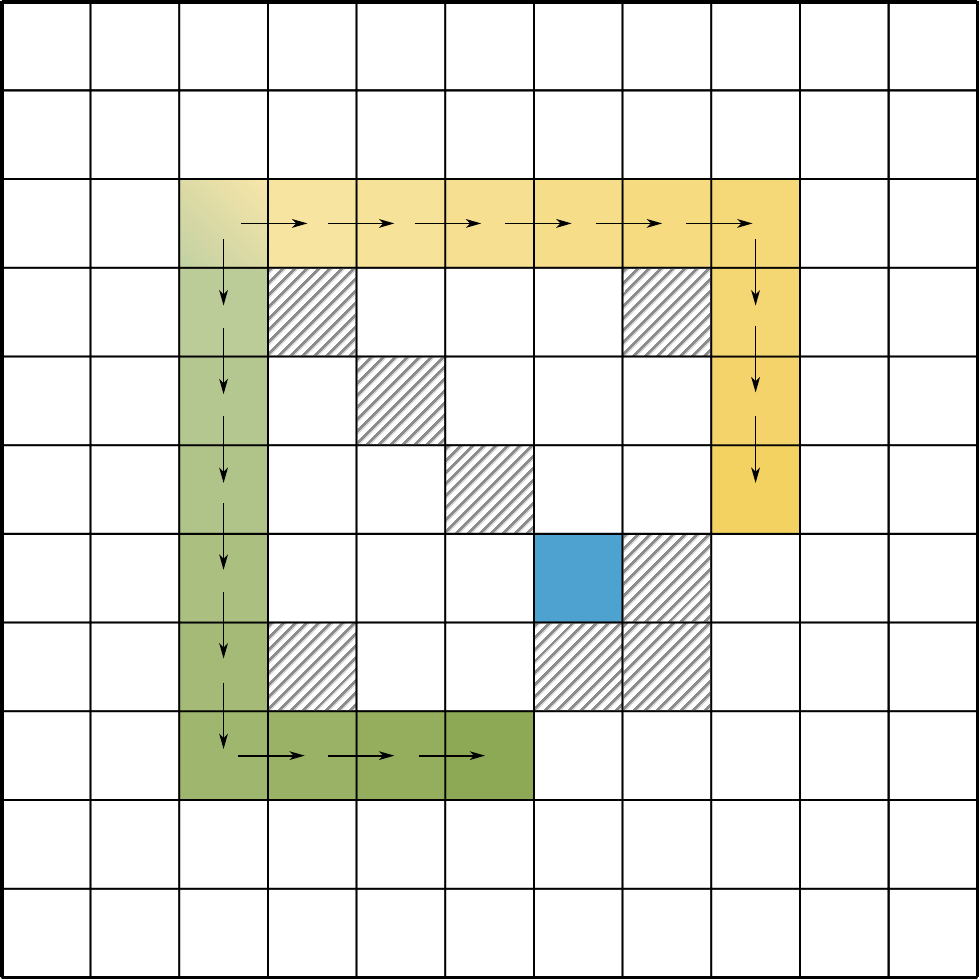}
    \caption{Graphical depiction of Gridworld layout~B, which is used for the changing reward path scenario.
    The blue cell represents the agent's starting position, gray-shaded cells are impassable for the agent.
    Green cells indicate the initial reward path, yellow cells represent the second reward path. 
    The direction of either reward path is indicated by arrows.
    }
    \label{fig:layout_b_vis}
\end{figure}

For the changing reward path scenario, or scenario B, we use a different Gridworld layout, which is presented in~\cref{fig:layout_b_vis} and hence named layout B.
Here, we have two reward paths: The initial reward path goes down and then right, with an episode length of nine.
After the switch happens, the second reward path is now active instead of the initial one.
This path is mirrored to the first path along the top-left to bottom-right diagonal.
The agent's starting position is on that diagonal, such that the number of rewarded action sequences is equal for both reward paths.
Further, in this layout, the sets of rewarded action sequences of the two reward paths are fully disjoint.
The key properties of both Gridworld layouts are summarized in~\cref{tab:layout_stats}.

\begin{table}[tb]
    \centering
    \caption{Statistics of Gridworld layouts~A and~B.}
    \label{tab:layout_stats}
        \begin{tabular}{L{2.5cm} R{1.5cm} R{1.5cm}}
            \toprule
            & Layout A & Layout B\\
            \midrule
            Episode Length & 7 & 7 \\[4pt]
            Length of Shortest & \multirow{2}{*}{4} & \multirow{2}{*}{3} \\
            Reward Path & &  \\[4pt]
            Initial Success & \multirow{2}{*}{0.017} & \multirow{2}{*}{0.041} \\
            Probability & & \\
            \bottomrule
        \end{tabular}
\end{table}

\subsubsection*{Stopping Criteria}
In each scenario, we test two different settings for deciding how many episodes an agent may train with each reward path.
These two stopping criteria allow comparing both the time required to reach a certain level of training as well as the performance after a fixed training time.

\begin{itemize}
    \item[\textbf{1.}] \textbf{``$k$ out of $n$'' successes}\\
    With this setting, we implement a switching or stopping criterion based on the agent's recent performance.
    If it manages to get $k$ rewards in the last $n$ tries, we consider the agent to have learned and switch to the second reward path or end the training, respectively.
    In this case, the training time is the most relevant metric, as the different agents can be expected to have similar learning progress upon triggering the ``$k$ out of $n$'' criterion.
    \item[\textbf{2.}] \textbf{Fixed number of episodes}\\
    Here, we fix the number of interaction episodes in advance as $N_1$ and $N_2$ for the initial and (optional) second reward path.
    With the same amount of training time, we are now more interested in the agents' final success probability.
\end{itemize}

\section{Results}
\label{sec:results}

\begin{table}[!b]
    \centering
    \caption{Learning times in terms of episodes for the hybrid and classical agent in the fixed reward path scenario with layout A and a ``4 out of 5'' stopping criterion (scenario A.1).
    The results stem from simulations with $N=1000$ runs per agent, the best value in each category is printed in boldface.}
    \label{tab:results_a1}
    \begin{tabular}{@{} L{2.5cm} *2{S[table-format=5.6,detect-weight,mode=text,uncertainty-mode=full]} @{}}
    \toprule
    & \multicolumn{2}{c}{Strategy} \\ 
    \cmidrule{2-3}
     & {Classical} & {Hybrid} \\
     \midrule
        First Reward & 58.4 \pm 1.8 & \B 17.9 \pm 0.4 \\
        \SI{20}{\percent} Success Prob. & 84.6 \pm 1.9 & \B 33.8 \pm 0.5 \\
        Completion & 94.2 \pm 1.9 & \B 44.5 \pm 0.5  \\
    \bottomrule
    \end{tabular}
\end{table}

\subsection*{Scenario A: Fixed Reward Path}
\label{sec:results:sub:fixed}
We combine scenario A with both the ```$k$ out of $n$' successes''-criterion, which we denote learning scenario A.1 and the the ''fixed number of episodes''-criterion, which we denote learning scenario A.2.

In the learning scenario A.1, we choose $k=4$ and $n=5$, such that training stops as soon as four out of the last five sampled action sequences are rewarded.
The comparison of the number of played episodes to reach certain key events for the hybrid and classical agent is shown in~\cref{tab:results_a1}.
To find the very first reward in a training run, the classical agent requires on average \num{58.4}~episodes.
With an average of \num{17.9}~episodes, the hybrid agent requires less than a third of the classical agent's learning time to reach this event.

Next, we have measured the average learning time required to reach a success probability of \SI{20}{\percent}, which was achieved in all runs.
Again, the hybrid agent requires, with \num{33.8}~episodes, less time than the classical agent's \num{84.6} episodes.
Lastly, we measure the average time each agent requires to complete the ``4 out of 5'' criterion.
Here, the hybrid agent still requires fewer episodes than the classical agent.
However, it can be noted that the difference in learning times is with \num{49.7} episodes roughly equal to the \num{50.8} episodes at the \SI{20}{\percent} success probability stage.

\begin{figure}[htb]
    \centering
    \includegraphics[width=\columnwidth]{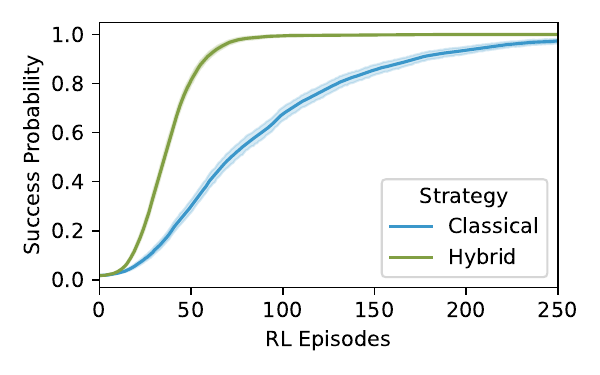}
    \caption{Comparison of the learning progress between the hybrid and the classical strategy for the fixed reward path scenario and $N_1=250$~training episodes (scenario A.2). 
    Average values of $N=500$~runs, error bands represent the \SI{95}{\percent} confidence interval.}
    \label{fig:results_a2}
\end{figure}

For learning scenario A.2 with a fixed number of interaction episodes, we set $N_1=250$ (and $N_2=0$, as we do not consider a switch of reward paths in scenario A).
We show the learning progress measured in terms of the agents' success probabilities in~\cref{fig:results_a2}.
The hybrid agent's learning curve shows a quicker initial increase and already after 100~episodes, it has reached an average success probability of~\SI{99.3}{\percent}. 
The classical agent reaches on average a success probability of~\SI{69.3}{\percent} after 100~episodes.
At the end of training after 250~episodes, the classical agent's success probability is~\SI{97.9}{\percent}, thus, roughly as large as the hybrid agent's success probability after 75~episodes.

\begin{table*}[!b]
    \centering
    \caption{Results for the hybrid and classical agent in the changing reward path scenario with layout B and a ``4 out of 5'' stopping criterion for both reward path phases (scenario B.1).
    Values indicate the mean with its standard error across $N=100$ runs per configuration.
    The lowest learning time of each category is printed in boldface.}
    \label{tab:results_b1}
    \begin{tabular}{@{} c p{2cm}  *3{S[table-format=3.6,detect-weight,mode=text,uncertainty-mode=full]} @{}}
    \toprule
      {$\gamma$} & Agent & {Episodes before Switch} & {Episodes after Switch} & {Episodes total} \\
    \midrule
      \multirow{2}{*}{\tablenum{0.01}} & Classical & 57.2 \pm 2.7 & 170.6 \pm 5.9 & 227.8 \pm 7.2 \\
      & Hybrid & \B 35.3 \pm 2.2 & 100.8 \pm 5.1 & 136.1 \pm 6.1 \\[3pt]
      \multirow{2}{*}{\tablenum{0.02}} & Classical & 60.5 \pm 3.5 & 126.7 \pm 4.2 & 187.2 \pm 5.6 \\
      & Hybrid & 40.6 \pm 1.6 & 84.1 \pm 3.4 & 124.7 \pm 4.1 \\[3pt]
     \multirow{2}{*}{\tablenum{0.05}} & Classical & 108.1 \pm 8.0 & 121.5 \pm 10.0 & 229.6 \pm 12.8 \\
      & Hybrid & 50.0 \pm 2.6 & \B 62.8 \pm 2.3 & \B 112.7 \pm 3.9 \\[3pt]
      \multirow{2}{*}{\tablenum{0.1}} & Classical & 371.1 \pm 32.5 & 380.8 \pm 33.8 & 751.9 \pm 44.3 \\
      & Hybrid & 80.0 \pm 4.9 & 85.1 \pm 5.1 & 165.1 \pm 6.6 \\
    \bottomrule
    \end{tabular}
\end{table*}

\subsection*{Scenario B: Changing Reward Path}
\label{sec:results:sub:changing}

In the changing reward scenario, we again distinguish between the two stopping criteria into two learning scenarios, B.1 and B.2.

In scenario B.1, we use the ```$k$ out of $n$' successes''-criterion with $k=4$ and $n=5$ both before and after the reward path switch.
An overview of the learning times for each reward path as well as the total learning time for this scenario is given in~\cref{tab:results_b1}.
For the first learning phase with the initial reward path, we can see that larger dissipation values consistently yield larger learning times for both agents.
The hybrid agent achieves lower learning times compared to the classical agent for fixed values of~$\gamma$.
After the switch to the second reward path, the hybrid agent still requires less learning time than the classical agent across all tested dissipation values. 
Here, however, an intermediate value of~$\gamma=0.05$ results in the lowest learning time among all configurations.
For the hybrid agent, this configuration is also the one with the lowest total number of episodes required in learning scenario B.1.
The classical agent, however, uses the least amount of episodes with the dissipation parameter set to~$\gamma=0.02$.

\begin{figure*}[!ht]
    \centering
    \includegraphics[width=\textwidth]{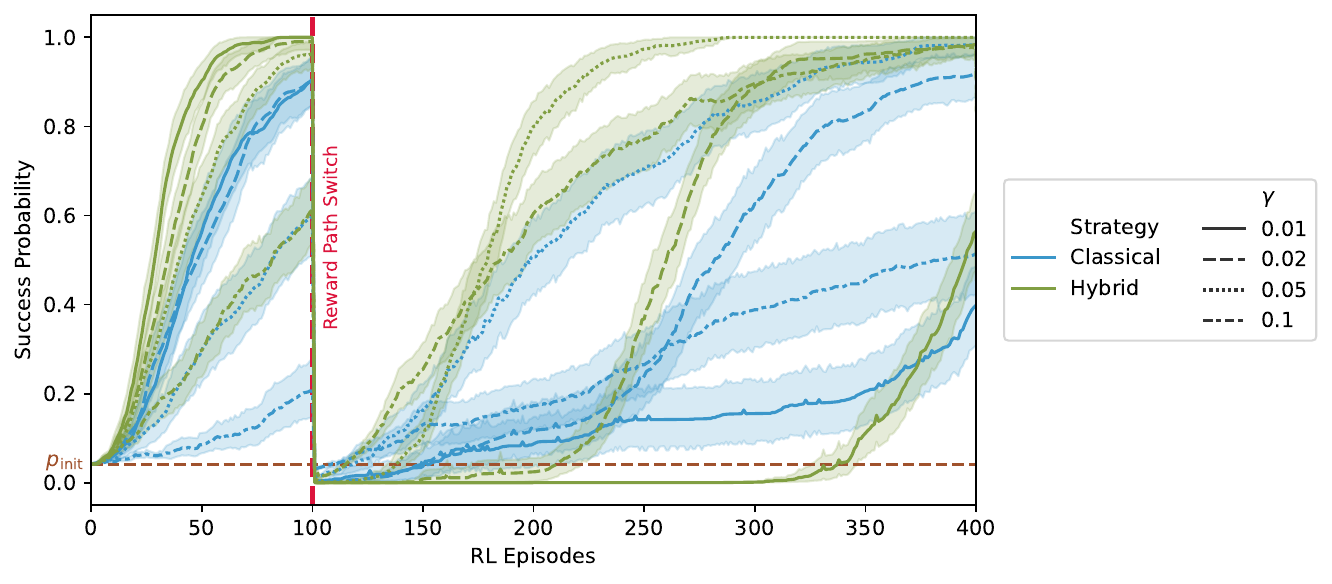}
    \caption{Comparison of the learning progress between the hybrid and the classical strategy for the changing reward path scenario with layout B, interaction times $N_1=100$ and $N_2=300$ (scenario B.2), and dissipation values $\gamma \in \{0.01, 0.02, 0.05, 0.1\}$.
    The dashed red vertical line indicates the change of reward paths after 100~episodes.
    The dotted horizontal line indicates the agents' initial success probability.
    Average values of $N=100$~runs, error bands represent the \SI{95}{\percent} confidence interval.}
    \label{fig:b2_learning_times}
\end{figure*}

For scenario B.2, we set the interaction times for the first and second reward paths to~$N_1=100$ and~$N_2=300$, respectively.
In~\cref{fig:b2_learning_times} we give an overview of the learning curves for the tested values of the dissipation parameter~$\gamma$.
A more detailed comparison of the two strategies depending on the dissipation value and an overview of the average success probabilities is given in~\cref{fig:b2_learning_curves_per_gamma} and~\cref{tab:b2_average_probs} of Appendix~\ref{app:sec:more-results}.
In the first learning phase, a quicker increase of the success probability for lower dissipation values can be noted for both strategies.
Comparing the two strategies, we observe quicker learning of the hybrid agent compared to the classical agent for fixed values of~$\gamma$.
In the second learning phase after the reward path switch, ranking the configurations is less straightforward.
For both the classical and the hybrid strategy, the learning curve for~$\gamma=0.01$ is the lowest throughout the complete 300 episodes of the second phase.
In terms of the average success probability in the second phase, the hybrid agent with $\gamma=0.05$ reaches the highest value with \SI{73.2}{\percent}.
This configuration also has the highest average success probability over the full 400~RL episodes.
At early stages after the reward path switch, however, the hybrid agent with $\gamma=0.1$ as well as the classical agents with $\gamma=0.05$ and $\gamma=0.1$ show a quicker initial increase in success probability.

\begin{figure}[t]
    \centering
    \includegraphics[width=\columnwidth]{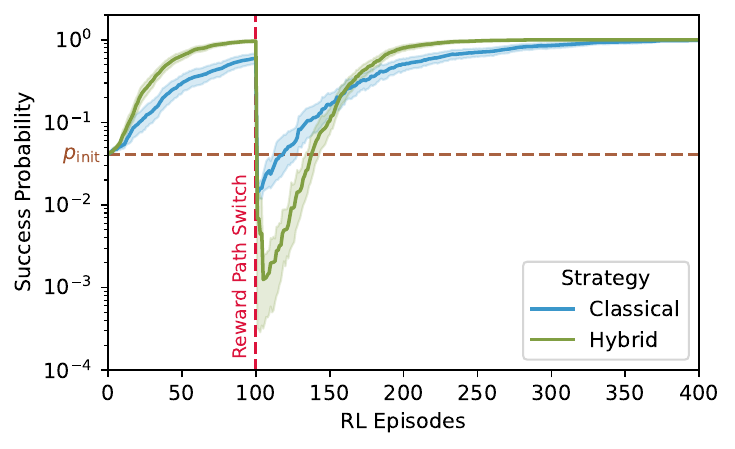}
    \caption{Comparison of the success probability for the classical and the hybrid agent with a dissipation value of~$\gamma=0.05$ on a logarithmic y-scale for the scenario B.2.
    Average values of $N=100$~runs, error bands represent the \SI{95}{\percent} confidence interval.}
    \label{fig:b2_best_gamma_log_y}
\end{figure}

In~\cref{fig:b2_best_gamma_log_y}, we illustrate the learning progress of the classical and the hybrid agent for $\gamma=0.05$ on a logarithmic y-scale.
The comparison shows that in the initial episodes following the reward path switch, the hybrid agent's success probability is lower by roughly one magnitude, before it increases sharply to surpass the classical agent after roughly 60~episodes on the new reward path.

\begin{figure}[t]
    \centering
    \includegraphics[width=\columnwidth]{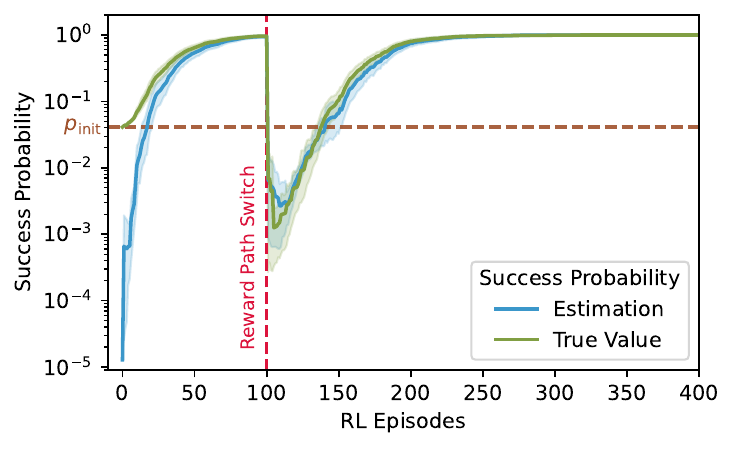}
    \caption{Comparison of the true success probability~$Q$ and the estimate $Q_\mathrm{est}$ during learning in scenario B.2 with the hybrid agent and a dissipation value of~$\gamma=0.05$.
    Average values of $N=100$~runs, error bands represent the \SI{95}{\percent} confidence interval.}
    \label{fig:b2_comparison_prob_true_est}
\end{figure}

Lastly, for the configuration with the hybrid agent and $\gamma=0.05$, we show the comparison between the true success probability, which we have looked at so far, and the agent's internal estimate $Q_\mathrm{est}$ in~\cref{fig:b2_comparison_prob_true_est}.
The estimate diverges most from the true success probability at the start of learning and after the reward path switch.
At the start of learning, the estimated success probability can be seen to be several orders of magnitude too low, before closing the gap as the agent learns.
After the reward path switch the estimated success probability is on average slightly larger than the true success probability for about \num{30}~episodes due to the incomplete purging of previously rewarded action sequences.
After this period, the estimated value is below the true value, matching the latter closely.

\section{Discussion}
\label{sec:discussion}

\subsection*{Scenario A: Fixed Reward Path}
Even though scenario A features a non-stationary reward function within the episode, the learning problem is similar to a purely stationary problem in the sense that the set of rewarded action sequences does not change.
The resulting learning curves of~\cref{fig:results_a2}, indicating quicker learning on average for the hybrid learning agent, are therefore in line with existing results~\citep{hamann2022performance}.
Further, the average classical time to find a first reward in scenario A.1 of $\langle t_\mathrm{class}\rangle = \num{58.4\pm1.8}$~episodes (cf.~\cref{tab:results_a1}) matches the  theoretical expected value of $58.7$~episodes well. 
This expected value follows from the idea that the classical agent's repeated sampling with the initial success probability $p_\mathrm{init}=0.017$ describes a geometric distribution, which terminates on average after $\frac{1}{p_\mathrm{init}}$ tries.
For the hybrid learning agent, the average time to find the first reward is with $\langle t_\mathrm{hybr}\rangle = \num{17.9\pm0.4}$~episodes even fairly below the theoretical upper bound~\citep{hamann2022performance} of $\frac{9}{2}\sqrt{\langle t_\mathrm{class}\rangle} = 34.4$~episodes.
This again is confirmation that the quadratic speedup in sample complexity persists in this scenario.
Lastly, the comparison of average learning times at different stages of the learning progress in~\cref{tab:results_a1} shows that the advantage in learning time is mainly gained at the early stages of learning, i.e., when the current success probabilities of the agents are low.
In fact, the gradual ramp-up of the possible number of AA iterations with successive non-rewarded outcomes (cf. lines~\ref{alg:hybrid-standard:line:k} and~\ref{alg:hybrid-standard:line:m} of~\cref{alg:hybrid-standard}) causes the hybrid agent to effectively converge to purely classical play in the limit of large success probabilities.

\subsection*{Scenario B: Changing Reward Path}
The results of scenario B indicate that the adaptations made to the hybrid learning agent enable it to function well in such continual RL problems.
Moreover, our results show that despite the larger success probability at the end of the first path and therefore a lesser chance for exploring new paths at the start of the second path phase, the hybrid learning agent eventually overtakes its classical counterpart in terms of learning progress in all tested configurations.

With respect to the dissipation mechanism, we note that a sensible choice of the dissipation value $\gamma$ is crucial to balance the trade-off between reaching a high success probability before the reward path switch and unlearning that preference quickly enough after the switch.
In the case of our problem layout B, a dissipation value of $\gamma=0.05$ proves best overall both in scenario B.1 (4 out of 5 successes) and in scenario B.2 ($N_1=100$, $N_2=300$).

Further, we note that the introduced adaptive procedure of computing $Q_\mathrm{est}$ matches the true success probability well throughout the learning process.
Especially after the reward path switch, $Q_\mathrm{est}$ overshoots the true value $Q$ only marginally, which is due to the efficient purging mechanism of previously rewarded action sequences.
Except these first episodes after the switch, $Q_\mathrm{est}$ still acts as a lower bound of $Q$, as intended in the original algorithm~\citep{hamann2022performance}.

Lastly, we emphasize that the chosen scenario is most extreme in the sense that the two sets of rewarded action sequences for the respective reward paths are disjoint.
In such a setting, good performance in the first learning phase is detrimental to the initial success chance in the second phase.
If a switch were to happen between two reward paths with an partially overlapping set of rewarded action sequences, the hybrid agent's better learning performance in the initial phase could instead contribute positively instead of negatively to subsequent phases, thus further increasing its advantage.

\section{Conclusion}
\label{sec:conclusion}
In this work we have conducted a first empirical investigation of the hybrid learning agent, which combines a Grover-type quadratic speedup in sample complexity with a classical RL algorithm, in non-stationary RL scenarios.
Using a modified Gridworld environment with different types of time-dependent reward functions, we compare the hybrid learning agent combined with the Projective Simulation (PS) RL algorithm to a purely classical PS agent with the same training hyperparameters.

Our results show that the hybrid agent's speedup naturally persists in the phase before the learning scenario requires any forgetting or relearning.
However, not least due to the minor modifications made to the hybrid agent introduced in this work, it is also capable to surpass the performance of its classical counterpart once a significant change in the environment has happened.
Given that the problem scenario in this article is chosen such that the hybrid agent's initial good performance puts it at a maximal disadvantage after a change in the reward function, we expect the results of this toy problem to generalize well to a wider range of problem cases.
Our findings motivate the conjecture that the speedup in learning achieved by our hybrid learning agent also enables it to better adapt its behavior to changing environments. 
This could also enable learning in fast changing environments where a corresponding classical agent would be too slow to learn at all.

The dissipation mechanism used in this paper in the policy update is a very simple mechanism to adapt the behavior of a learning agent to changing environments. 
Here, the agent does not actively monitor possible changes in the environment, but adapts passively to them instead. 
This is a good method for making the learning of an agent more robust to, e.g., initial non-optimal behavior as well as small changes in the environment. 
If disruptive changes of the environment are expected, more proactive methods for adapting the learning of the hybrid agent to these changes can be beneficial. 
Therefore, further investigations include broadening the scope of non-stationarity to, e.g., time dependency in the transition probability function.
In addition, the combination of the hybrid learning agent with algorithms dedicated to continual RL which, unlike the PS algorithm used here, prevent catastrophic forgetting, is another interesting direction for continued research.

\backmatter

\bmhead{Acknowledgements}
This article was written as part of the Qu-Gov project, which was commissioned by the German Federal Ministry of Finance. 
M.~R. and L.~S. want to extend their gratitude to Kim Nguyen, Manfred Paeschke, Oliver Muth, Andreas Wilke, and Yvonne Ripke for their continuous encouragement and support.
O.~S. and S.~W. thank Annette Zapf for helpful comments.

\section*{Declarations}

\bmhead{Competing interests}
The authors declare no competing interests.


\begin{appendices}

\renewcommand{\thefigure}{\arabic{figure}}
\setcounter{figure}{6}

\renewcommand{\thetable}{\arabic{table}}
\setcounter{table}{3}

\section{Additional Plots and Tables for Scenario B.2}
\label{app:sec:more-results}
\begin{table}[htb]
    \centering
    \caption{Average success probabilities of the hybrid and the classical strategy for the changing reward path scenario with interaction times $N_1=100$ and $N_2=300$ (scenario B.2) and dissipation values $\gamma \in \{0.01, 0.02, 0.05, 0.1\}$.
    Values indicate the mean with its standard error across $N=100$ runs per configuration.
    The highest average success probability of each category is printed in boldface.}
    \label{tab:b2_average_probs}
    \begin{tabular}{@{}c m{1.3cm}  *3{S[table-format=2.5,detect-weight,mode=text]} @{}}
    \toprule
      \multirow{3}{*}{$\gamma$} & \multirow{3}{*}{Agent} & \multicolumn{3}{c}{Avg. Success Probability [\%]}\\
       \cmidrule{3-5}
       &  & {Before} & {After} & {\multirow{2}{*}{Total}} \\
       &  & {Switch} & {Switch} &  \\
    \midrule
      \multirow{2}{*}{\tablenum{0.01}} & Classical & 50.0 \pm 0.4 & 13.5 \pm 0.2 & 22.7 \pm 0.2 \\
       & Hybrid & \B 70.3 \pm 0.4 & 5.5 \pm 0.1 & 21.6 \pm 0.2  \\[3pt]
      \multirow{2}{*}{\tablenum{0.02}} & Classical & 49.7 \pm 0.4 & 39.2 \pm 0.3 & 41.8 \pm 0.2 \\
       & Hybrid & 64.2 \pm 0.4 & 45.8 \pm 0.3 & 50.5 \pm 0.2 \\[3pt]
      \multirow{2}{*}{\tablenum{0.05}} & Classical & 29.4 \pm 0.4 & 61.6 \pm 0.3 & 53.5 \pm 0.2 \\
       & Hybrid & 56.8 \pm 0.4 & \B 73.2 \pm 0.2 & \B 69.0 \pm 0.2 \\[3pt]
      \multirow{2}{*}{\tablenum{0.1}} & Classical & 11.2 \pm 0.3 & 28.8 \pm 0.3 & 24.4 \pm 0.2 \\
       & Hybrid & 30.7 \pm 0.3 & 66.4 \pm 0.2 & 57.4 \pm 0.2 \\
    \bottomrule
    \end{tabular}
\end{table}

\begin{figure*}[htb]
    \centering
    \includegraphics[width=0.8\textwidth]{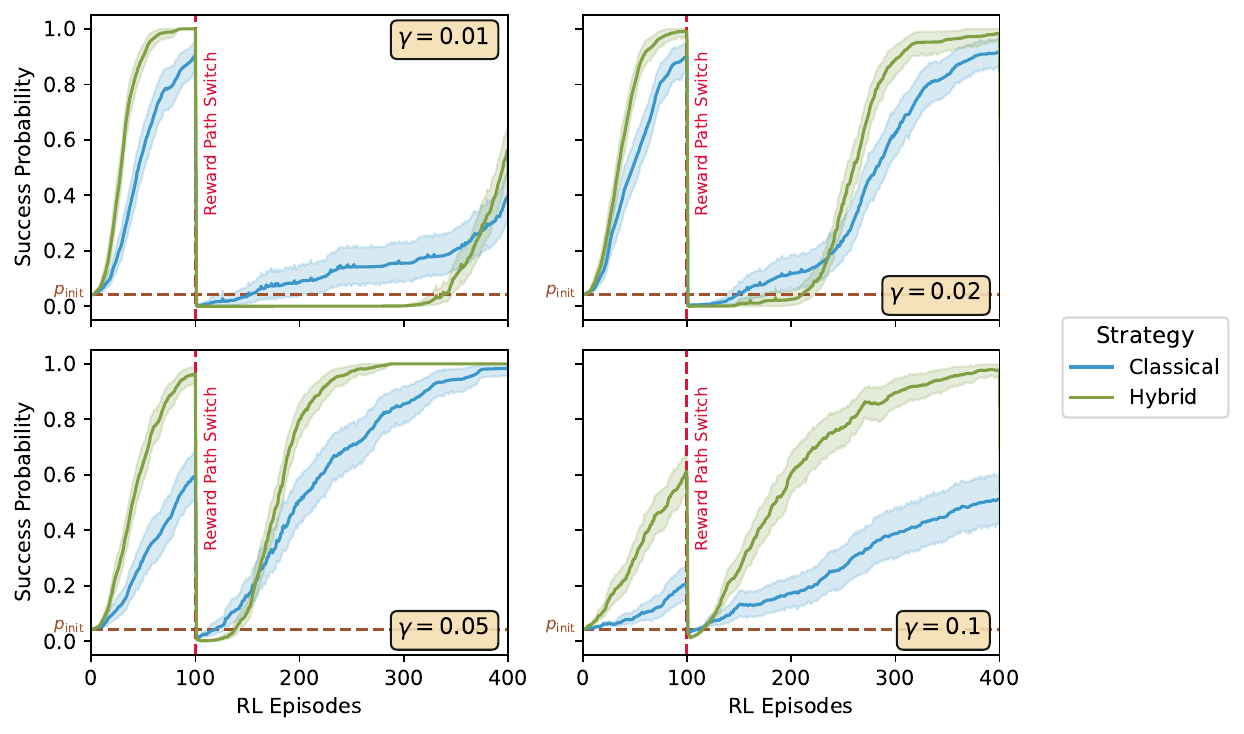}
    \caption{Learning curves of the hybrid and the classical strategy for the changing reward path scenario B.2 with interaction times $N_1=100$ and $N_2=300$ and dissipation values $\gamma \in \{0.01, 0.02, 0.05, 0.1\}$.
    The dashed red vertical line indicates the change of reward paths after 100~episodes.
    The dotted horizontal line indicates the agents' initial success probability.
    Average values of $N=100$~runs, error bands represent the \SI{95}{\percent} confidence interval.
    }
    \label{fig:b2_learning_curves_per_gamma}
\end{figure*}

\section{Projective Simulation}
\label{app:sec:projective-simulation}

Projective Simulation~\citep{briegel2012projective} is a (tabular) RL method based on a random walk along a learned network of nodes or \textit{clips}, with transition probabilities governed by a clip's outgoing edge weights.

\begin{figure}[htb]
    \centering
    \includegraphics[width=0.95\columnwidth]{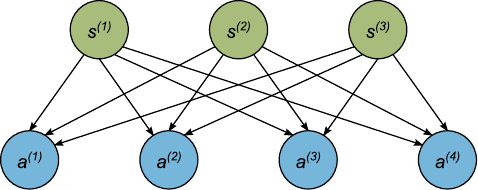}
    \caption{Illustration of the basic ECM layout with exemplary percept and action sets $\mathcal{S}=\{\, s^{(1)}, \, s^{(2)}, \, s^{(3)}\, \}$ and $\mathcal{A}=\{ \, a^{(1)},\, a^{(2)},\, a^{(3)},\, a^{(4)}\, \}$, respectively.
    Arrows from the percept (green) to the action (blue) clips indicate possible transitions, with an \textit{h-value} assigned to each of them.
    }
    \label{fig:basic_ecm}
\end{figure}

A clip may either be an action $a$ or a percept $s$.
Through interaction, the agent discovers connections from action to percept clips and vice versa.
These connections are stored in form of a graph network called the \textit{episodic \& compositional memory} (ECM), which builds up over the process of learning.
In an MDP, this network has the form of one layer of percept clips, which is fully connected to a layer of action clips as illustrated in~\cref{fig:basic_ecm}.
An input percept triggers a random walk starting at that respective percept clip and results in outputting the action associated to the clip at which the random walk terminates.

Edges from a percept $s$ to an available action $a$ are assigned a so-called \textit{h-value} $h(s,a)$, which is initially set to one.
The agent's policy $\pi$ is determined by either linearly weighing the h-values 
\begin{align}\label{eq:linear-policy}
    \pi(a|s) = \cfrac{h(s,a)}{\sum\limits_{a\in\mathcal{A}}h(s,a)}
\end{align}
or by using the \textit{softmax} function
\begin{align}\label{eq:softmax-policy}
    \pi(a|s) = \cfrac{e^{\beta \cdot h(s,a)}}{\sum\limits_{a\in\mathcal{A}}e^{\beta \cdot h(s,a)}}
\end{align}
with some pseudo-temperature parameter $\beta\geq0\in\mathbb{R}$, which was introduced by~\citep{melnikov2018benchmarking}.

After an interaction step, the h-values $h_{t+1}(s,a)$ of the subsequent time step $t+1$ are updated by adding the received reward $r_t(s,a)$ to the h-values of the current time step $t$:
\begin{align}\label{eq:basic-update}
    h_{t+1}(s,a) = h_{t}(s,a) + r_t(s,a)
\end{align}

To this basic PS framework, several additions have been made, of which we mention the ones relevant to our problem.
First, the \textit{dissipation} mechanism introduced in the original paper~\citep{briegel2012projective} enables the agent to forget learned preferences, resulting in its adaptability to changing environments.
With the introduction of the dissipation parameter $\gamma\in[0,1]$, we add a term to the basic h-value update rule of~\cref{eq:basic-update}:
\begin{align}\label{eq:dissipation-update}
    h_{t+1}(s,a) = h_{t}(s,a) + r_t(s,a) - \gamma \big( h_{t}(s,a) - 1 \big)
\end{align}
The effect of this term for values $\gamma \in (0,1)$ is to drive all h-values to some degree towards their initial value of 1 after each interaction step.
The case $\gamma = 0$ obviously retrieves the scenario without dissipation, whereas the case $\gamma=1$ simplifies~\cref{eq:dissipation-update} to $h_{t}(s,a) = 1 + r_t(s,a)$ and thus essentially resets the agent to its initial h-values after each time step.

Second, we present the \textit{edge glow} or \textit{afterglow} mechanism which was introduced in~\citep{mautner2015projective}.
The purpose of this mechanism is to help the agent deal with delayed rewards in episodic environments.
In many RL environments which are games, e.g., chess or black-jack, non-zero rewards are not issued during the game, but just upon the ultimate outcome.
Thus, if such an episode consists of $L$ time steps, $r_t(s,a) \neq 0$ just for $t=L$ and the combination of percept and action encountered/played in that time step.
In the following, we use the shorthand notation $r$ to denote this reward value at the end of the episode.
Naturally, not just the final action in the episode, but potentially the entire sequence of preceding actions has an impact on the outcome of the episode.

The idea of edge glow is to attribute a contribution of previous actions to receiving a delayed reward, with the contribution being more significant the closer the action was temporally to the reward.
To this end, this concept introduces the notion of \textit{excitations} of edges in the ECM with the so-called glow value $g$.
Initially, the glow value is set to zero for all transitions in the ECM, $g_0(s,a)=0\;\; \forall s\in\mathcal{S}, a\in\mathcal{A}$.
An edge corresponding to selecting action $a$ at percept $s$ is excited when this $(s,a)$ combination occurs in the interaction.
Upon excitation at time $t$, the glow value is set to one for this edge, $g_t(s,a)=1$.
With time, these excitations decay towards zero following the update rule
\begin{align}\label{eq:glow-update}
    g_{t+1}(s,a) = \left( 1 - \eta  \right) \cdot g_{t}(s,a)  \quad \forall s\in\mathcal{S}, a\in\mathcal{A}
\end{align}
with glow decay parameter $\eta\in[0,1]$.
Upon updating the h-values, the glow value modulates the reward as
\begin{align}\label{eq:h-update-with-glow}
    h_{t+1}(s,a) =h_{t}(s,a) + g_{t}(s,a) \cdot r - \gamma \big( h_{t}(s,a) - 1 \big).
\end{align}
As actions played in a previous episode are not meaningful to a current episode's outcome, all glow values must be reset to zero at the start of a new episode in episodic environments.

In this work, we use PS solely for episodic RL tasks with a binary reward at the end of an episode.
It is therefore evident that updates of the h-values at intermediate steps within an episode have just a contribution coming from the dissipation term.
To enable quicker computation, we omit this dissipation update within the episode in our implementation such that we perform just one update at the end of each episode.
Note that works which apply dissipation at each time step likely require lower values for the dissipation parameter due to the increased frequency of dissipation updates.
The full procedure to update the h-values and the policy in PS using dissipation, glow, and a softmax policy is shown in pseudocode form in~\cref{alg:ps-policy-update}.

\algrenewcommand\algorithmicensure{\textbf{Input:}}
\begin{algorithm}[!tb]
\caption{Projective Simulation version of \texttt{policy\_update}}\label{alg:ps-policy-update}
\begin{algorithmic}[1]
\Ensure sequence of played actions \\$\Vec{a}=(a_0, a_1, \dots, a_{L-1})$, \\
sequence of encountered percepts \\$\Vec{s}=(s_0, s_1, \dots, s_L)$, \\
boolean reward information $\texttt{rewarded}$
\Require table of h-values $h(s,a)$ for all actions $a\in\mathcal{A}$ and percepts $s\in\mathcal{S}$, \\dissipation parameter $\gamma$, \\glow parameter $\eta$, \\softmax temperature parameter $\beta$

\State initialize glow values \\$g(s,a)=0\;\; \forall s\in\mathcal{S}, a\in\mathcal{A}$
\If{\texttt{rewarded}}
    \State $r \gets 1$
    \For{$i=0$ \textbf{to} $L-1$}
        \LComment{glow decay}
        \State $g(s, a) \gets (1-\eta) \cdot g(s,a) \;\; \forall s\in\mathcal{S}, a\in\mathcal{A}$ 
        \LComment{glow reset for excitation} 
        \State $g(s_i, a_i) \gets 1$ 
    \EndFor
\Else
    \State $r \gets 0$
\EndIf
\For{$(s,a) \in \mathcal{S}\otimes \mathcal{A}$}
    \LComment{h-value update}
    \State $h(s,a) \gets h(s,a) + g(s,a)\cdot r - \gamma \big( h(s,a) - 1 \big)$ 
    \LComment{policy computation}
    \State $\pi(a|s) \gets \cfrac{e^{\beta \cdot h(s,a)}}{\sum\limits_{a\in\mathcal{A}}e^{\beta \cdot h(s,a)}}$ 
\EndFor
\end{algorithmic}
\end{algorithm}

Finally, the so-called \textit{mapping} focuses on the compositional aspect of the ECM.
Here, the agent builds the ECM network as a map of the environment to learn the connectivity between different states.
This is contrary to the two-layer input-output coupling shown in~\cref{fig:basic_ecm}.
Instead, the node corresponding to a chosen action is connected to the node representing the next percept.
If that percept has not yet been part of the learning history, a new node is created instead.
This way, using its past knowledge, the agent can anticipate to which state an action may lead given its current state.
In deterministic environments especially, an agent may thus not just compute the probability $\pi(a|s)$ according to its policy to execute action $a$ in state $s$.
Instead, it can also compute the probability to execute a sequence of actions $\Vec{a}$ as it has knowledge of the intermediate states these actions will lead to (if the map is not built up yet for a certain part of the action sequence, the uniform probability is again assumed).
This possibility to compute a full action sequence probability $\hat{\pi}(\Vec{a}|s)$ (we omit the condition on the state $s$ in the main text as we always start from a fixed initial state $s_0$) is crucial for the hybrid learning agent.

\subsection{Hybrid Agent: repeated diffusion}
\label{app:sec:projective-simulation:sub:repeated-diffusion}

To accommodate for the fact that the hybrid learning agent updates less frequently compared to a classical RL agent due to the overhead of interaction episodes required for amplitude amplification, we must increase the strength of the dissipation per update to enable an equal ``forgetting'' rate.
One method to achieve this would be a $2k$-fold application of the dissipation update for each of the $2k$ episodes spent with amplitude amplification before the update with the actual reward information.
This, however, appears largely inefficient.
Our goal, therefore, is to adapt the dissipation part of the update rule of~\cref{eq:dissipation-update} for the hybrid agent such that the two following updates are equal:
\begin{itemize}
    \item one update after $2k+1$ episodes of using the hybrid agent with no rewards during episodes 1 to $2k$ and a possible reward in episode $2k+1$,
    \item continuous updates in $2k+1$ episodes of classical play with no reward during episodes 1 to $2k$ and a possible reward in episode $2k+1$.
\end{itemize}
In this case, as described above, we update just at the end of an episode instead of after every time step, we use the uppercase index $h^{(\tau)}(s,a)$ to denote the h-value of state-action pair $(s,a)$ throughout episode $\tau$ and in the same manner $r^{(\tau)}$ to denote the binary reward received in episode $\tau$.
In this notation, the update rule of PS with dissipation of~\cref{eq:dissipation-update} reads:
\begin{equation}\label{eq:dissipation-update-episode}
    h^{(\tau+1)}(s,a) = h^{(\tau)}(s, a) + r^{(\tau)} - \gamma\big(h^{(\tau)}(s,a) - 1\big).
\end{equation}
We claim that the update rule
\begin{align}\label{eq:h-update-multiple-episodes}
    h^{(t + N)} = \big(h^{(t)} - 1 \big) \cdot \left( 1 - \gamma \right)^{N} + 1 + r^{(t+N-1)}
\end{align}
is equivalent to apply $N-1$ updates with just dissipation and no reward, $r^{(\tau)}=0$ for $\tau \in \{t,\dots,t+N-2\}$, and one update with dissipation and a possible reward $r^{(t+N-1)}$ and, therefore, fulfills the goal defined above.
This equivalence can be shown via induction as follows (to simplify notation, we abbreviate $h^{(\tau)}(s,a)$ as $h^{(\tau)}$).
For $N=1$, some reshuffling of terms yields
\begin{align*}
    h^{(t+1)} \:&\stackrel{\mathclap{(\ref{eq:dissipation-update-episode})}}{=}\: h^{(t)} + r^{(t)} - \gamma(h^{(t)} - 1)  \\
        &= \big(h^{(t)} - 1 \big) \cdot \left( 1 - \gamma \right)^{1} + 1 + r^{(t)}.
\end{align*}
For the induction step $N\rightarrow N+1$, we have 
\begin{align*}
    &h^{(t+N+1)} \\
    \stackrel{\mathclap{(\ref{eq:dissipation-update-episode})}}{=}\;& h^{(t+N)} - \gamma(h^{(t+N)} - 1) + r^{(t+N)} \\
    =\;& h^{(t+N)} (1 - \gamma) + \gamma + r^{(t+N)}\\
    \stackrel{\mathclap{(\ref{eq:h-update-multiple-episodes})}}{=}\;& \left[\big(h^{(t)} - 1 \big) \cdot \left( 1 - \gamma \right)^{N} + 1 + r^{(t+N-1)}\right] \left( 1- \gamma \right) \\
    &+ \gamma + r^{(t+N)} \\[4pt]
    =\;& \big(h^{(t)} - 1 \big) \cdot ( 1 - \gamma )^{(N + 1)} + (1 - \gamma) \\
    &+ r^{(t+N-1)}(1-\gamma) + \gamma + r^{(t+N)} \\[4pt]
    =\;& \big(h^{(t)} - 1 \big) \cdot ( 1 - \gamma )^{(N + 1)} + 1 + r^{(t+N)},
\end{align*}
where in the last equality we have used the fact that intermediate rewards are zero. This completes the proof.

\section{Estimation of the Probability Lower Bound}
\label{app:sec:probability-estimation}
As we have mentioned in our description of scenario A, the hybrid agent as presented in Algorithm~\ref{alg:hybrid-standard} can be used without any adaptations.
However, we need to specify how we initialize (line~\ref{alg:hybrid-standard:line:q_min-init} of Algorithm~\ref{alg:hybrid-standard}) and later update (line~\ref{alg:hybrid-standard:line:q_min-update} of Algorithm~\ref{alg:hybrid-standard}) the lower bound estimate $Q_\mathrm{min}$ of the current success probability.

As the initial estimate, we choose $Q_\mathrm{min}=|\mathcal{A}|^{-T}$, which corresponds to one rewarded path among the $|\mathcal{A}|^T$ possible paths for the chosen action set $\mathcal{A}$ and episode length $T$.
Here, we must ensure that the chosen episode length is sufficient to yield at least one rewarded path.
Further, we assume that each path is sampled with equal probability initially.
In tabular versions of Q-Learning and SARSA with an equal initialization of q-values and an epsilon-greedy policy with random tie-breaks as well as in Projective Simulation, this assumption holds true.

As soon as we have found a rewarded action sequence, we can improve the estimate.
We introduce the initially empty set $\mathcal{R}_\mathrm{found}$ of rewarded action sequences the agent has encountered during its learning.
If the agent now samples an action sequence $\Vec{a}$ which is rewarded, we first investigate how many of the $T$ actions were necessary to reach the target.
Doing so, we can split the action sequence $\Vec{a} = \Vec{a}_\mathrm{trunc} \oplus \Vec{a}_\mathrm{rest}$ into the truncated part of length $T_\mathrm{trunc} \leq T$ and a rest part $\Vec{a}_\mathrm{rest}$ of length $ T_\mathrm{rest}\geq 0$, $T_\mathrm{trunc} + T_\mathrm{rest} = T$ such that executing $\Vec{a}_\mathrm{trunc}$ exactly reaches the target.
Then, we add the truncated action sequence to $\mathcal{R}_\mathrm{found}$.
After a first reward is found, we estimate the lower bound for $Q$ as:
\begin{align}\label{eq:lower-bound-estimate}
    Q_\mathrm{min} = \sum\limits_{\Vec{b}\in \mathcal{R}_\mathrm{found}} \hat{\pi}(\Vec{b})\,,
\end{align}
which is still a lower bound by standard probabilistic arguments. 
Furthermore, to ensure that $Q_\mathrm{min}$ is strictly larger than zero, we must require the policy to fulfill $\pi(a|s) > 0 \; \forall a\in\mathcal{A}, s\in\mathcal{S}$.

The chosen procedure usually results quite quickly in ``good" lower bounds $Q_{\text{min}}$ close to the actual probability $Q$. 
However, it is also possible to use predetermined hard-coded lower limits. 
In this case, the effort to remember all rewarded action sequences and adapt them as described in the next section is not necessary.

In general, the hybrid learning agent is relatively robust to lower bounds $Q_\text{min}$ which are either very low or even wrong because the value is too high. 
In the case of a very low bound, ramping up the maximal number of Grover iterations and randomly sampling  the number of amplitude amplification rounds $k$ (see~\Cref{alg:hybrid-standard}) solves the problem that the resulting probability of finding a rewarded sequence of action decreases again if $k$ is chosen too large.
If the estimation of $Q_\text{min}$ is too large (such as in our example problem after the environment has changed) the hybrid agent will perform on average less steps $k$ of amplitude amplification. 
In this case, the hybrid agent still achieves a speedup in learning. 
However, no quadratic scaling can be achieved, as discussed in detail in \citep{hamann2022performance}.

\subsection{Adaptation for Changing Reward Paths}
\label{app:sec:probability-estimation:sub:adaption}
\begin{algorithm}[!tb]
\caption{Adapted Hybrid Learning Agent}\label{alg:hybrid-adapted}
\begin{algorithmic}[1]
    \vspace{3pt}
    \Require MDP $\langle \,T, \,\mathcal{S}, \,\mathcal{A}, \,p_t^{(\mathrm{det})}(s,a), \,r_t(s,a) \,\rangle$, \\initial percept $s_0$, \\Grover operator $G$, \\$\texttt{policy\_update}$ subroutine
    \State initialize algorithm constant $\lambda \gets 5/4$ 
    \State initialize set of found rewards $\mathcal{R}_\mathrm{found} \gets \emptyset$
    \State initialize lower bound estimate $Q_\mathrm{min} \gets |\mathcal{A}|^{-T}$
    \State algorithm parameter $m \gets 1$ \label{alg:hybrid-adapted:line:loop-start}
    \State $\texttt{rewarded} \gets \texttt{false}$\label{alg:hybrid-adapted:line:continue}
    \LComment{Quantum Part}
    \State $k \gets \mathrm{random \;integer \; in \;[0,m)}$
    \State prepare $\ket{\psi} \gets \sum\limits_{\Vec{a}\in \mathcal{A}^{\otimes T}} \sqrt{\hat{\pi}(\Vec{a})} \ket{\Vec{a}}_A \ket{\vec{s}_\mathrm{init}}_S \ket{-}_R$
    \State amplitude amplification $\ket{\psi'} \gets G^k \ket{\psi}$ 
    \State $\Vec{a}'\gets$ \textbf{measure}  $\ket{\psi'}$
    \LComment{Classical Part}
    \For{$i=1$ \textbf{to} $T$}
        \State $s_i \gets p_i^{(\mathrm{det})}(s_{i-1}, a_{i-1})$
        \State $r_i \gets r_i(s_{i-1},a_{i-1})$
        \If{$r_i = 1$}
            \State \texttt{rewarded} $\gets$ \texttt{true}
            \State $\Vec{s} \gets (s_0, s_1, \dots, s_i)$
            \State $\Vec{a}'_\mathrm{trunc} \gets (a_0', a_1', \dots, a_{i-1}')^\top$
            \State $\mathcal{R}_\mathrm{found} \gets \mathcal{R}_\mathrm{found} \cup \{\Vec{a}'_\mathrm{trunc}\}$
        \EndIf
    \EndFor
    \LComment{Classical RL Update}
    \State update $\pi$ using subroutine $\texttt{policy\_update}(\Vec{a}', \Vec{s}, \texttt{rewarded})$
    \LComment{Update of Probability Estimate}
    \If{\texttt{not rewarded}}
        \State purge $\vec{a}'$ (or a possible truncation $\Vec{a}'_\mathrm{trunc}$) from $\mathcal{R}_\mathrm{found}$ if $\vec{a}' \in \mathcal{R}_\mathrm{found}$
    \EndIf
    \If{$|\mathcal{R}_\mathrm{found}| \geq 1$}
        \State $Q_\mathrm{min} \gets \sum\limits_{\Vec{a}\in \mathcal{R}_\mathrm{found}} \hat{\pi}(\Vec{a})$
    \Else
        \State $Q_\mathrm{min} \gets |\mathcal{A}|^{-T}$
    \EndIf
    \If{\texttt{rewarded}}
        \State repeat from step~\ref{alg:hybrid-adapted:line:loop-start} 
    \Else
        \State $m \gets \min\left( \lambda \cdot m, \: Q_\mathrm{min}^{-1/2} \right)$
        \State repeat from step~\ref{alg:hybrid-adapted:line:continue}   
    \EndIf
    
\end{algorithmic}
\end{algorithm}
To enable the hybrid learning agent for scenario B, which introduces a switch to a different reward path after some time during training, we must ensure two aspects: 
On the one hand, a forgetting mechanism is crucial to unlearn the preference for the initial path.
In this regard, the hybrid agent does not differ from classical RL algorithms.
However, this forgetting mechanism is typically implemented in the RL algorithm the hybrid agent is used in combination with.
On the other hand, the estimation method of the current success probability must be adapted, which is unique to the hybrid agent. 
While the forgetting mechanism lowers occurrence probabilities for the previously rewarded action sequences over time, as they are no longer reinforced, this happens slowly and does not correct for the fact that these action sequences should no longer contribute to the estimate~$Q_\mathrm{min}$.

To continue with the estimation method as in~\cref{eq:lower-bound-estimate} in the changing reward path scenario, we must remove action sequences from $\mathcal{R}_\mathrm{found}$ if we find them to be no longer rewarded.
This \textit{purging} procedure can be implemented as follows.
Whenever we sample an action sequence $\Vec{a}$ which is not rewarded, we compare it to all action sequences of~$\mathcal{R}_\mathrm{found}$.
If the sequence~$\Vec{a}$ or some possible truncation~$\Vec{a}_\mathrm{trunc}$ is an element of $\mathcal{R}_\mathrm{found}$, it is no longer rewarded after the switch and hence dropped from the set $\mathcal{R}_\mathrm{found}$.

The obvious implication of this procedure is that the estimate built from summing the occurrence probabilities of all action sequences in $\mathcal{R}_\mathrm{found}$ is no longer a lower bound of the true success probability.
To avoid ambiguity in~\cref{alg:hybrid-adapted}, which describes the adapted hybrid agent for the changing reward path scenario, we rename the estimate $Q_\mathrm{min}$ as $Q_\mathrm{est}$.
Whether the hybrid agent can perform well in the post-switch phase with this new estimation scheme is a key question in our experiments.
Furthermore, it may occur that all previously rewarded action sequences are purged from~$\mathcal{R}_\mathrm{found}$ before a new rewarded sequence is found.
In this case, a fall-back option to estimate $Q_\mathrm{est}$ is to use the lowest occurrence probability of any action sequence.
This however, requires computing the occurrence probability of all action sequences -- which has an exponential computational cost with respect to the episode length -- because the the algorithm has learned in the meantime and the starting estimate of $| \mathcal{A}|^{-T}$ likely no longer holds.

\end{appendices}

\end{document}